\documentclass{an}
\usepackage{graphicx}
\usepackage{times}
\begin{document}

% The following seven commands are intended for editorial usage and
% should be ignored by the author(s).
\Pagespan{1}{}% Document's page range. 
% If second parameter is left empty, the last page is computed
% automatically.
\Yearpublication{2015}%
\Yearsubmission{2015}%
\Month{1}%   
\Volume{999}%  
\Issue{92}% 
% \DOI{This.is/not.aDOI}% 

\title{GPS/CSS radio sources and their relation to other AGN}

\author{Elaine\,M.\,Sadler\inst{1}\fnmsep\thanks{Corresponding author:
  \email{elaine.sadler@sydney.edu.au}}
% Example for footnote, note the usage of the \texttt{fnmsep} command
% as separator between institute number and footnote mark}
}
\titlerunning{GPS/CSS sources and other AGN}
\authorrunning{Elaine\,M.\,Sadler}
\institute{
Sydney Institute for Astronomy, School of Physics, University of Sydney, NSW 2006, Australia; \\
ARC Centre of Excellence for All-sky Astrophysics (CAASTRO)  }

\received{XXXX}
\accepted{XXXX}
\publonline{XXXX}


\abstract{%
We are entering a new era of sensitive, large-area and multi-frequency radio surveys that will allow us to identify Gigahertz-Peaked Spectrum (GPS) and Compact Steep Spectrum (CSS) radio sources over a wide range in radio luminosity and study them within the context of the overall radio-source populations to which they belong. 
`Classical' GPS/CSS objects are extremely luminous radio sources with a compact double morphology, commonly thought to represent the earliest stages in the life cycle of powerful radio galaxies (e.g. O'Dea 1998). It is now becoming easier to identify GPS/CSS candidates with much lower radio luminosity -- particularly in the nearby Universe. These less powerful objects, with typical 1.4\,GHz radio luminosities of $10^{23}$ to $10^{25}$\,W\,Hz$^{-1}$, include peaked-spectrum radio sources with a core--jet morphology on parsec scales as well as high-frequency GPS-like peaked components embedded within lower-frequency extended emission. In the latter case, the presence of a young GPS component may not be evident from low-frequency data alone. Many radio galaxies in the local Universe have a compact (FR-0) morphology, and appear to lack extended radio emission on kiloparsec scales. The relationship of these FR-0 objects to the classical GPS/CSS radio sources remains unclear -- some of them may represent short-lived episodes of AGN activity that will not lead to an extended FR-1 or FR-2 radio galaxy. Future wide-band radio surveys will shed more light on this -- such surveys should ideally be coordinated to cover the full frequency range from 100\,MHz to 100\,GHz in order to sample all stages of GPS/CSS evolution in an unbiased way. 
}
\maketitle

\section{Introduction}
Gigahertz-Peaked Spectrum (GPS) and Compact Steep Spectrum (CSS) sources are compact, powerful radio sources with well-defined peaks in their radio spectra (near 1\,GHz for GPS sources, and 100\,MHz for CSS objects). They are estimated to make up around 10\% (GPS) and 30\% (CSS) of the bright radio-source population (O'Dea 1998). 

\begin{figure}
\includegraphics[width=0.9\linewidth]{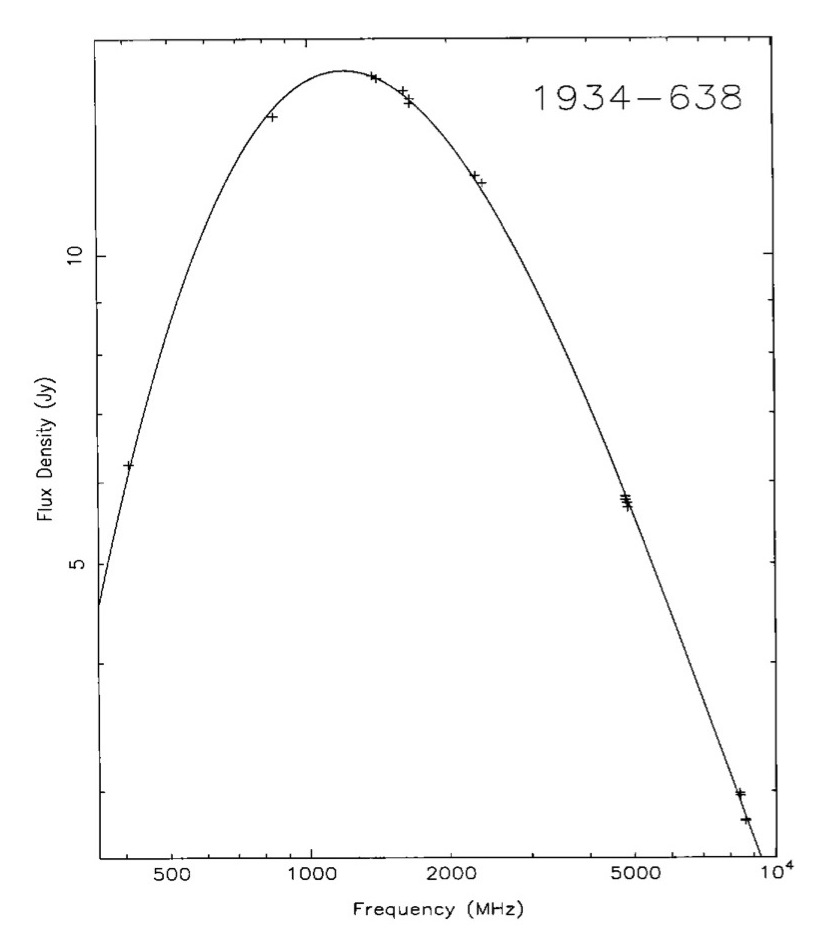}
\caption{Radio spectrum of the canonical GPS source PKS\,1934--638 between 300\,MHz and 8\,GHz  (Reynolds 1994). This source, associated with a galaxy at redshift $z=0.18$, has a spectral peak just above 1\,GHz. PKS\,1934--638 is used as the primary flux calibrator for the Australia Telescope Compact Array, so has been observed regularly over many years. }
\label{fig_1934_spec}
\end{figure}

\begin{figure}
\includegraphics[width=0.8\linewidth]{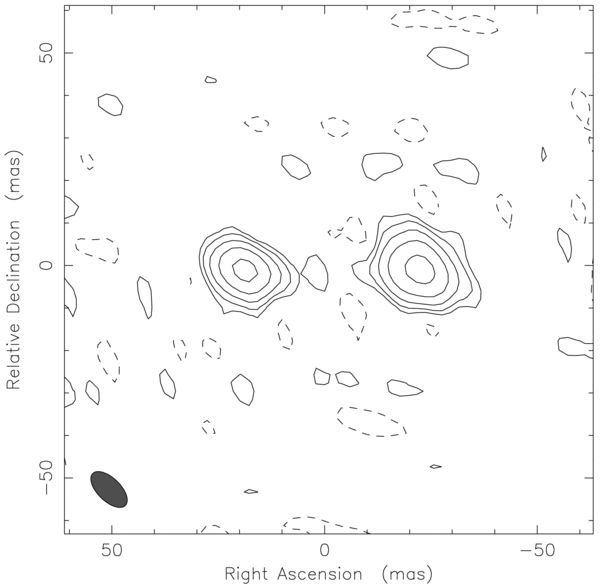}
\caption{VLBI image of the GPS source PKS\,1934--638 at 1.4\,GHz, showing the compact double structure of the radio emission (Tzioumis et al.\,2010).  }
\label{fig_1934_vlbi}
\end{figure}

Figure \ref{fig_1934_spec}\ shows the radio spectrum of the $z=0.18$ radio galaxy PKS\,1934-638 (Fosbury et al.\ 1987; Reynolds 1994), often taken as a prototype for the GPS class (O'Dea, Baum \& Stanghellini 1991). With a 1.4\,GHz radio luminosity of 10$^{27.1}$ W\,Hz$^{-1}$, this is one of the most powerful radio galaxies in the nearby Universe. 
The compact double structure of the source can be seen in the VLBI image in Figure \ref{fig_1934_vlbi}. The two hotspots are separated by 42\,mas (Tzioumis et al.\ 2010), corresponding to a projected linear separation of around 130\,pc. 

In this paper, I will discuss the relationship between GPS and CSS sources and the overall population of radio AGN at the same redshift, with a particular focus on nearby (redshift $z<0.2$) objects. Throughout this paper I adopt a standard cosmological model with  H$_{\rm 0}$ = 71 km\,s$^{-1}$\,Mpc$^{-1}$, $\Omega_{\rm M}$ = 0.27 and $\Omega_{\lambda}$ = 0.73. 

\subsection{Radio-source populations in the local Universe} 
Over the past decade, our understanding of the overall radio-source population in the nearby Universe has advanced substantially, mainly through the analysis of large and complete radio-source samples assembled by combining large-area radio continuum and optical redshift surveys (e.g. Sadler et al.\ 2002; Best et al.\ 2005; Mauch \& Sadler 2007; Best \& Heckman 2012; van Velzen et al.\ 2012). A recent review of the key results is given by Heckman \& Best (2014). 

\begin{figure}
\includegraphics[width=\linewidth]{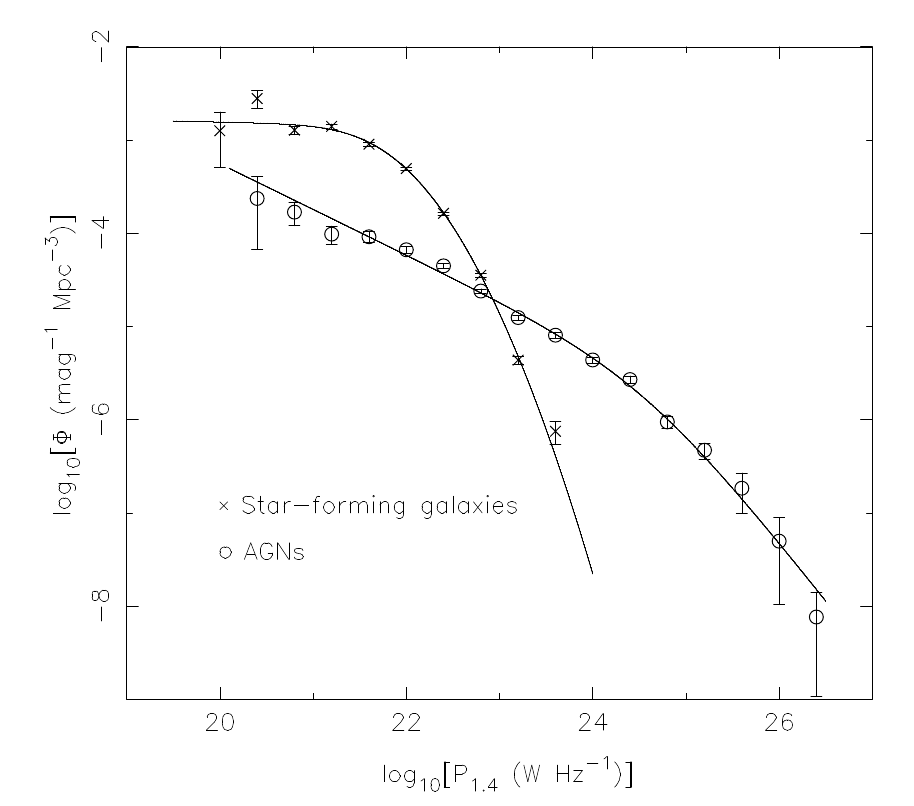}
\caption{The radio luminosity function at 1.4GHz derived by Mauch \& Sadler (2007) for radio-loud AGN (circles) and star-forming galaxies (crosses) in the local Universe (median redshift $z=0.05$), measured by cross-matching the 6dF Galaxy Survey (6dFGS; Jones et al.\ 2009) with the NVSS radio catalogue (Condon et al.\ 1998).  }
\label{fig_rlf}
\end{figure}

The local radio luminosity function has now been measured over more than six orders of magnitude in radio power, as shown in Figure\,\ref{fig_rlf}. Optical spectra usually allow low-power radio AGN to be distinguished from `normal' galaxies whose radio luminosity is dominated by  processes related to star formation (Condon 1989), and there is a large range in radio power below 10$^{24}$ W\,Hz$^{-1}$ where both radio AGN and star-forming galaxies are common.  

\subsection{Two populations of radio AGN} 
\begin{figure}
\includegraphics[width=\linewidth]{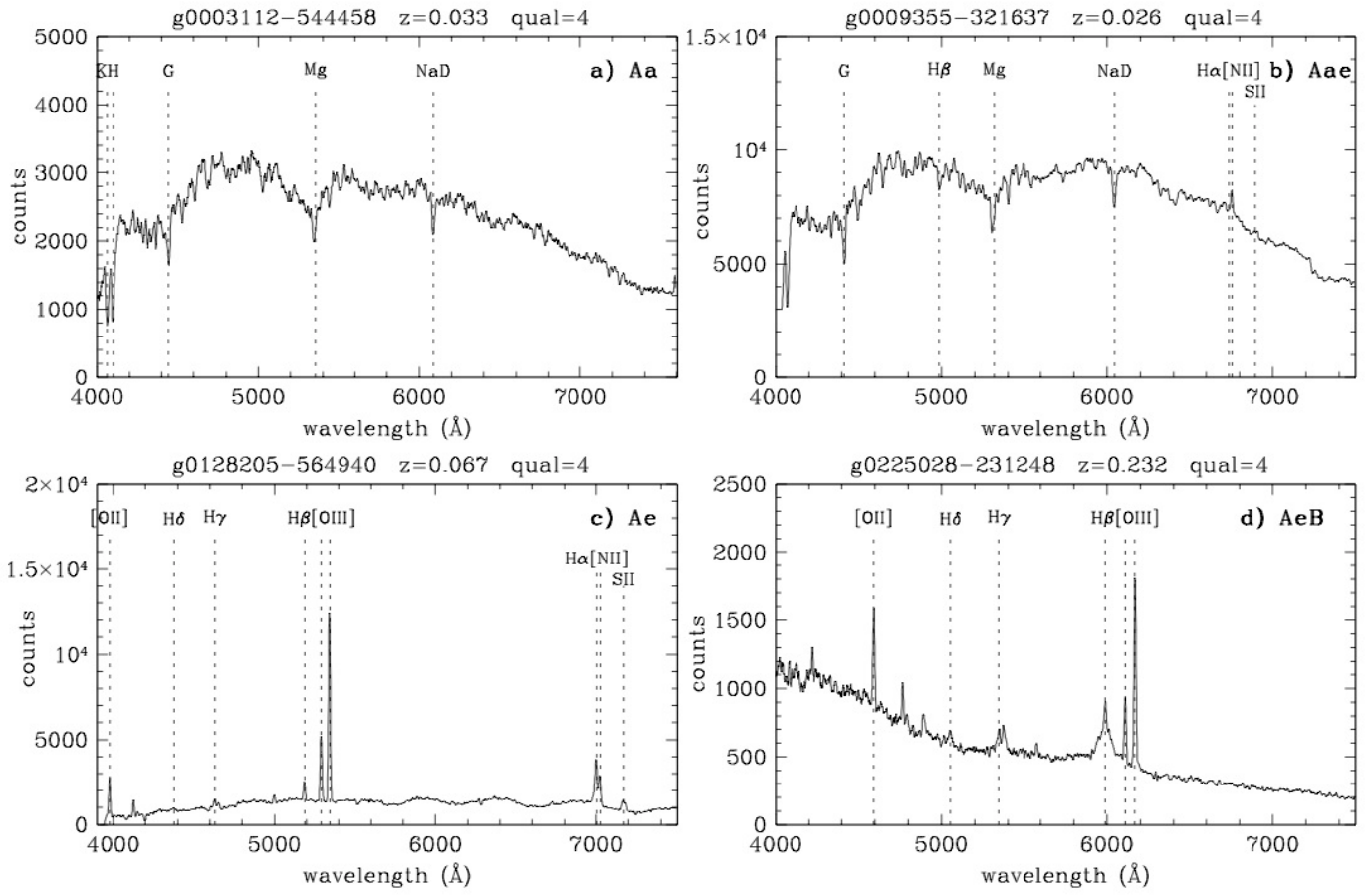}
\caption{6dFGS optical  spectra of two nearby galaxies with radio AGN detected in the AT20G survey:  {\it (top)} the low-excitation radio galaxy (LERG) PKS\,0000-550 at redshift $z=0.033$ and {\it (bottom)} the high-excitation radio galaxy (HERG) PMN\,J0128-5649 at $z=0.067$.  Note that the LERG spectrum shows no optical signature of AGN activity.    }
\label{fig_spectra}
\end{figure}

Over the past few years, several authors (e.g. Hardcastle et al. 2007, Best \& Heckman 2012 and references therein) have proposed that there is a fundamental dichotomy between Ôhigh-excit\-ation radio galaxiesÕ (HERGs), in which the AGN is fuelled in a radiatively efficient way by a classic accretion disk, and Ôlow-excitation radio galaxiesÕ (LERGs) in which the accretion rate is significantly lower and accretion is radiatively inefficient. Best \& Heckman (2012) derive accretion rates of one and ten per cent of the Eddington rate for HERGs, in contrast to a typical accretion rate below one per cent Eddington for the LERGs in their sample. Once again the two classes can usually be distinguished spectroscopically, as illustrated in Figure\,\ref{fig_spectra}.

{\it High-excitation radio galaxies}\ (HERGs) have a classical accretion disk surrounding the central black hole, which gives rise to strong, high-excitation optical emission lines. Unified models, in which a central dusty torus obscures the central broad line region at some orientations (e.g. Urry \& Padovani 1995) are expected to apply to HERGs, which may appear as either radio-loud QSOs with broad Balmer lines or narrow-line radio galaxies depending on the viewing angle. 

{\it Low-excitation radio galaxies}\ (LERGs) lack this classical accretion disk, and as a result their optical spectra show weak or no emission lines even though the central black hole can power radio jets. In these objects, which make up the vast majority of radio AGN in the nearby Universe, accretion onto the black hole is radiatively inefficient. Since there is no broad-line region and (probably) no dusty torus surrounding the black hole, the unified models developed for `classical' AGN do not apply to LERGs. 

While the most powerful local radio galaxies (those with radio luminosity above 10$^{26}$ W\,Hz$^{-1}$ at 1.4\,GHz) are usually HERGs, Best \& Heckman (2012) measured the local radio luminosity function separately for HERGs and LERGs and showed that both populations co-exist over a wide range in radio power. There is also no simple relationship between radio morphology and the HERG/LERG classification. Most FR-1 radio galaxies have low-excitation optical spectra, but the FR-2 radio galaxies are a mixture of high-excitation and low-excitation objects. 

\subsection{Identifying GPS and CSS sources} 
\begin{figure}
\hspace*{-0.4cm}
\includegraphics[width=1.05\linewidth]{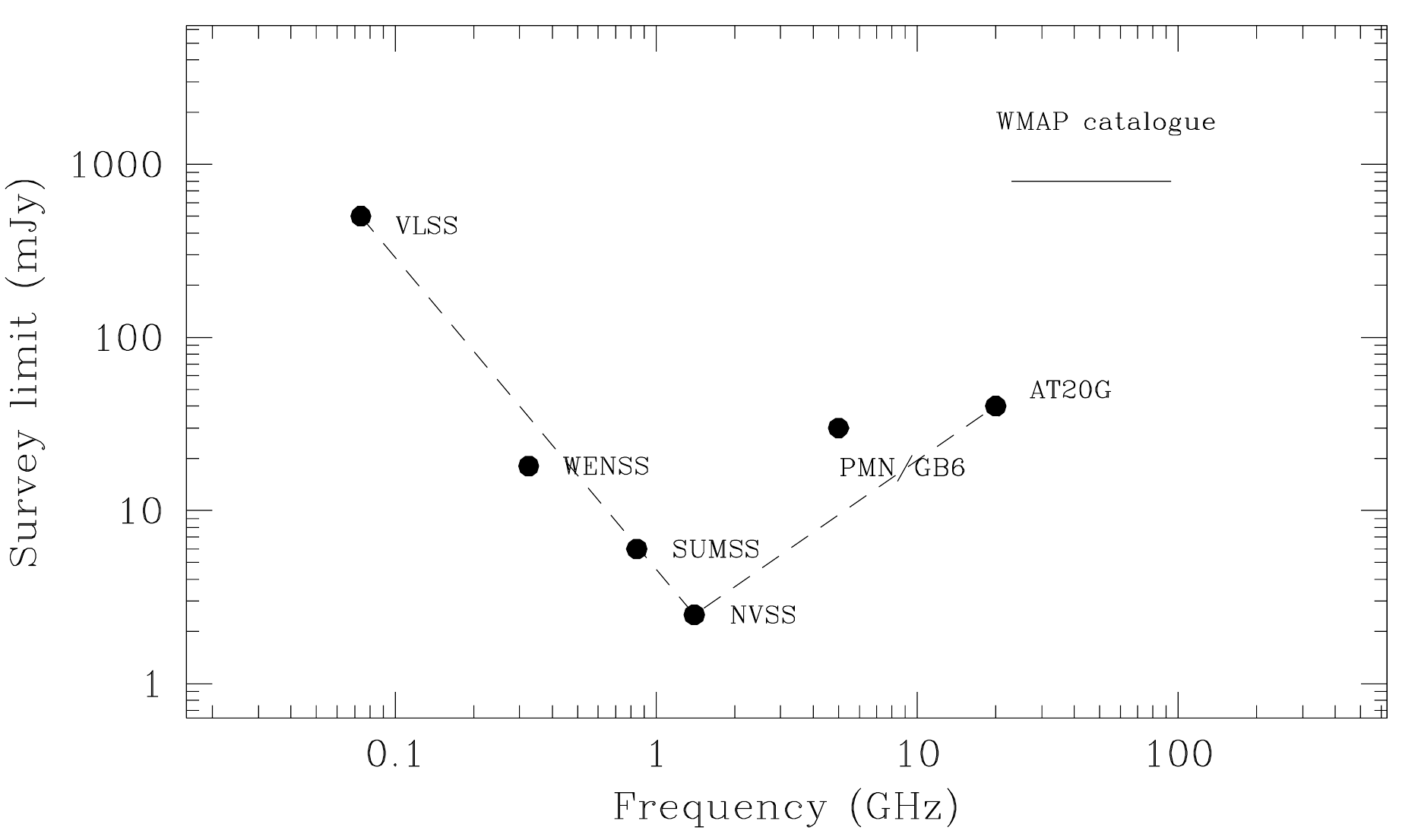}
\caption{Detection limits of some large-area ($>10,000$\,deg$^2$) radio surveys at different frequencies (adapted from Sadler et al.\ 2008). These include the VLA Low-Frequency Sky Survey (VLSS; Cohen et al.\ 2007) at 74\, MHz, the Westerbork Northern Sky Survey (WENSS; Rengelink et al.\ 1997) at 325\,MHz, the Sydney University Molonglo Sky Survey (SUMSS; Mauch et al.\ 2003) at 843\,MHz, the NRAO VLA Sky Survey (NVSS; Condon et al. 1998) at 1.4\,GHz, the PMN/GB6 surveys at 4.85\,GHz (Gregory et al.\ 1994, 1996) and the Australia Telescope 20\,GHz survey (AT20G; Murphy et al.\ 2010). 
 }
\label{fig_surveys}
\end{figure}

GPS and CSS sources cannot be identified from a single-frequency radio survey because of their peaked radio spectra, so finding them usually requires combining the data from several different surveys. As Figure\,\ref{fig_surveys}\ shows, the sensitivity of current large-area radio surveys is highest at 1.4\,GHz and drops off significantly at both higher and lower frequencies. 

As has long been recognized (e.g. O'Dea et al.\ 1991), this has two consequences.  First of all, sources peaking near 1\,GHz will be easier to recognize than those that peak at higher or lower frequencies (though for low-frequency peakers this will change in the near future with the availability of new and more sensitive low-frequency surveys from the MWA and LOFAR telescopes, as discussed in the presentations by Callingham and Mahony at this workshop). 

Secondly, it means that existing lists of GPS and CSS radio sources (O'Dea et al.\ 1991; Labiano et al.\ 2007; de Vries et al.\ 2007; Randall et al.\ 2011) mainly contain brighter (and therefore more luminous) GPS and CSS sources because these are easier to identify using currently-available surveys. For example, the GPS sources listed by O'Dea (1998) have a median radio luminosity above $10^{27}$\,W\,Hz$^{-1}$ at 5\,GHz, and  even the sample of `low-luminosity compact sources' studied by Kunert-Bajrazewska et al.\ (2010) mainly contains sources with radio luminosities between $10^{25}$\,W\,Hz$^{-1}$ and $10^{26}$\,W\,Hz$^{-1}$ at 1.4\,GHz. 

As a result, the well-studied GPS/CSS sources,with 1.4 GHz radio luminosities above $10^{25}$\,W\,Hz$^{-1}$, 
are much bright\-er than the bulk of the local radio-source population shown in Figure\,\ref{fig_rlf}. This leads to the two main questions I'll address in this paper: \\
\begin{itemize}
\item 
Is there a large population of  lower-luminosity (radio power  $<10^{25}$\,W\,Hz$^{-1}$) GPS/CSS sources? 
\item
If so, how do their properties compare to the more luminous GPS and CSS samples studied to date? 
\end{itemize}
 
\begin{figure}
\includegraphics[width=\linewidth]{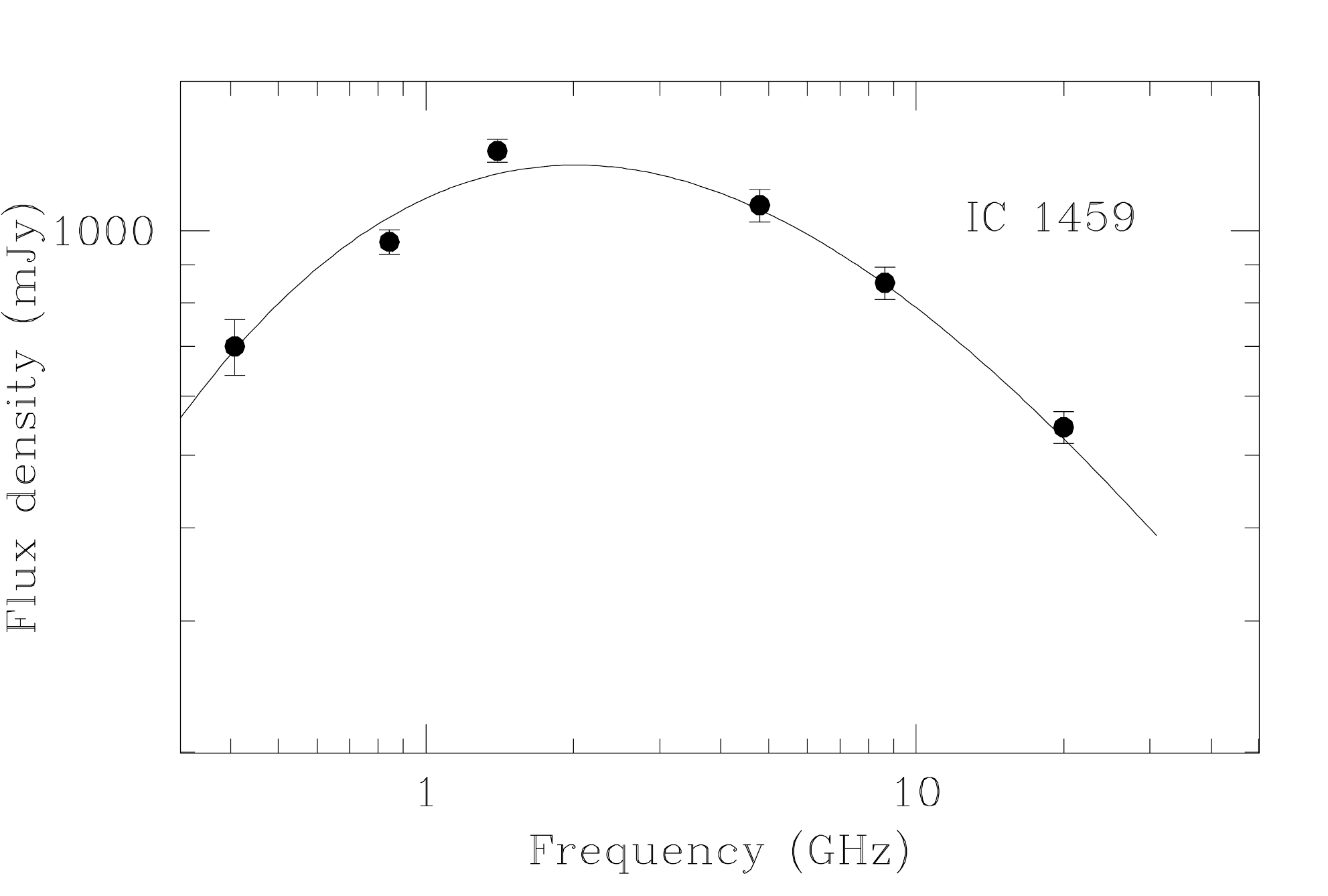}
\caption{Radio spectrum of the nearby, low-luminosity radio galaxy IC\,1459. This galaxy was first recognized as a GPS source by Tingay et al.\ (2003), and belongs to the AT20G-6dFGS galaxy sample discussed in \S2.}
\label{fig_ic1459_spec}
\end{figure}

\subsection{Lower-luminosity GPS and CSS radio sources} 
There is already good evidence for the existence of lower-power GPS sources, including some with radio luminosities well below 10$^{24}$\,W\,Hz$^{-1}$. 
Figure\,\ref{fig_ic1459_spec}\ shows the radio spectrum of the galaxy IC\,1459 at $z=0.005$, which has been identified by Tingay et al.\ (2003) as one of the closest GPS radio sources.  IC\,1459 has a radio spectrum peaking near 2\,GHz, and a 1.4\,GHz radio luminosity of $10^{23.0}$\,W Hz$^{-1}$. 

\begin{figure}
\hspace*{1.0cm}
\includegraphics[width=0.8\linewidth]{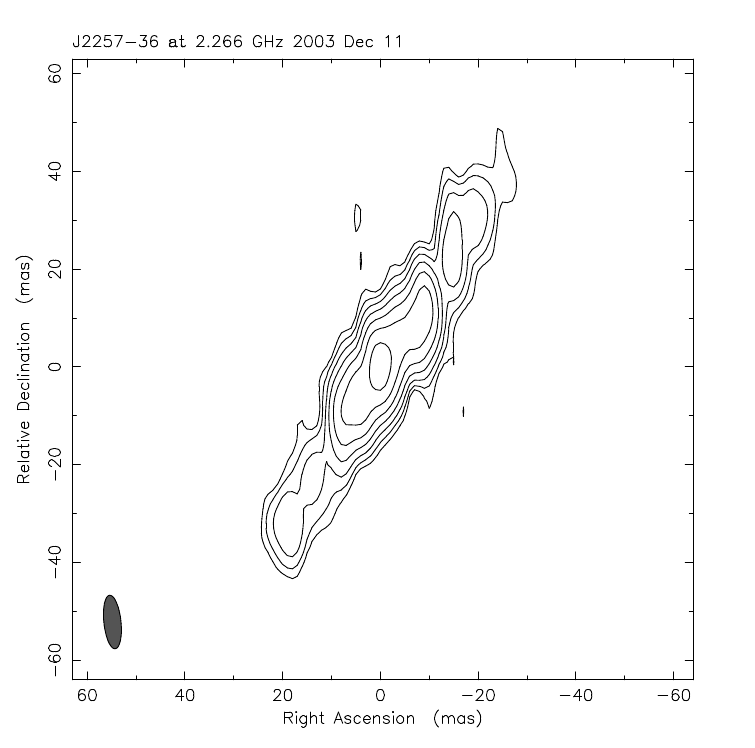}
\caption{VLBI image of the nearby radio source IC\,1459, from Tingay \& Edwards (2015). This low-luminosity GPS source shows a core-jet structure with a projected linear size of around 10\,pc, rather than the compact double structure seen in more powerful objects like PKS\,1934-638 shown in  Figure\,\ref{fig_1934_vlbi}. }
\label{fig_ic1459_vlbi}
\end{figure}

As can be seen from Figure\,\ref{fig_ic1459_vlbi}, this source has a core-jet morphology on VLBI scales, rather than the compact double structure usually seen in more powerful GPS sources like PKS\,1834-638 (see Figure\,\ref{fig_1934_vlbi}). 
Tingay et al.\ (2015) have proposed that there is a luminosity-dependent break in the morphology of GPS radio galaxies, analogous to the FR-1/FR-2 luminosity-morph\-ology break seen in extended radio galaxies (Fanaroff \& Riley 1974), and that IC\,1459 belongs to a sub-class of faint GPS sources with a core-jet structure which may be  the progenitors of FR-1 radio galaxies. 
These authors also point out that low-luminosity GPS sources are likely to be strongly under-represented in current GPS samples, because even at distances as low as 40--50\,Mpc their radio emission is faint enough to drop below the $\sim1$\,Jy flux density limit used for many previous GPS/ CSS searches. 

The ability to identify and study complete samples of GPS radio sources in the nearby Universe is an important step in developing and testing models that can explain both the unique characteristics of these sources and their evolution with time (e.g. Snellen et al.\ 2003; Tingay et al.\ 2003). Ideally, we would like to have simultaneous, multi-frequency radio data for a large sample of galaxies spanning only a small range in redshift (so that any effects of redshift evolution are removed). If possible, this should be combined with radio measurements at several epochs in time (to characterize variability and identify any beamed sources), along with VLBI imaging of the small-scale radio structure. 
Such data are now starting to become available, as major radio telescopes around the world are equipped with wide-band receivers and correlators that make it possible to carry out multi-frequency surveys more rapidly and repeat them at several epochs. 

\section{A local galaxy sample with multi-frequency radio data}

To select candidate GPS/CSS sources from within a well-defined sample of radio AGN, we use data from the Australia Telescope 20\,GHz (AT20G; Murphy et al.\ 2010) survey, which provides near-simultaneous measurements at 5, 8, and 20\,GHz for a sample of over 5,000 radio sources selected at 20\,GHz.  

This catalogue has already been used to select samples of radio sources with spectral peaks above 5\,GHz (e.g. Hancock 2009). Here, we select a volume-limited sub-sample of nearby AT20G sources for which good-quality optical spectra are also available.

\subsection{Sample selection} 
Sadler et al.\ (2014) have recently cross-matched the AT20G data with the 6dF Galaxy Survey (6dFGS; Jones et al.\ 2009) to produce a volume-limited sample of 202 high-frequency radio sources associated with galaxies in the local Universe.   

Of the 202 sources in the AT20G-6dFGS sample, 201 are identified with radio AGN (the only exception is the nearby starburst galaxy NGC\,253).  Coupled with the multi-frequency AT20G data, the AT20G-6dFGS dataset allows us to identify candidate GPS and CSS sources from within a well-defined sample of radio AGN that are all at similar redshift. 
The median redshift of galaxies in the AT20G-6dFGS sample is $z=0.058$, and the median radio luminosity at 1.4\,GHz is $3.2\times10^{24}$\,W\,Hz$^{-1}$. The low-luminosity radio regime is well-sampled, since 55 of the galaxies in the sample have a 1.4\,GHz radio luminosity below $10^{24}$\,W\,Hz$^{-1}$.  

The 6dFGS optical spectra indicate that about  a quarter (23\%) of the AT20G-6dFGS sample are high-excitation radio AGN (HERGs), while the great majority (77\%) are low-excitation systems (LERGs). The HERG fraction is however significantly higher in this sample than in similar samples of radio AGN selected at lower radio frequencies (e.g. Best \& Heckman 2012). 

\subsection{Source structure at 20\,GHz} 
The AT20G snapshot images typically have an angular resolution of 10--15\,arcsec (Murphy et al. 2010), corresponding to a projected linear size of 10--15 kpc for galaxies at $z\sim0.06$\  (the median redshift of galaxies in our sample). In most cases, therefore, a source that is unresolved in the 20\,GHz images is confined within its host galaxy. This size scale is characteristic of Compact Steep Spectrum (CSS) radio sources, which are usually smaller than 15\,kpc in extent (Fanti et al.\ 1990). 

Information is also available at higher resolution from analysis of the AT20G data on the longest (6\,km) ATCA baseline, as discussed by Chhetri et al.\ (2013). These authors used data from the longest (6\,km) ATCA baseline to determine how much of the radio emission seen by the AT20G survey arose in very compact components, and showed that AT20G sources with flat radio spectra ($\alpha^{20}_{1} > -0.5$) at 1--20\,GHz generally had almost all their 20\,GHz radio emission arising from a central source less than about 0.2\,arcsec in angular size. 

\begin{figure}
%\hspace*{1.0cm}
\includegraphics[width=0.95\linewidth]{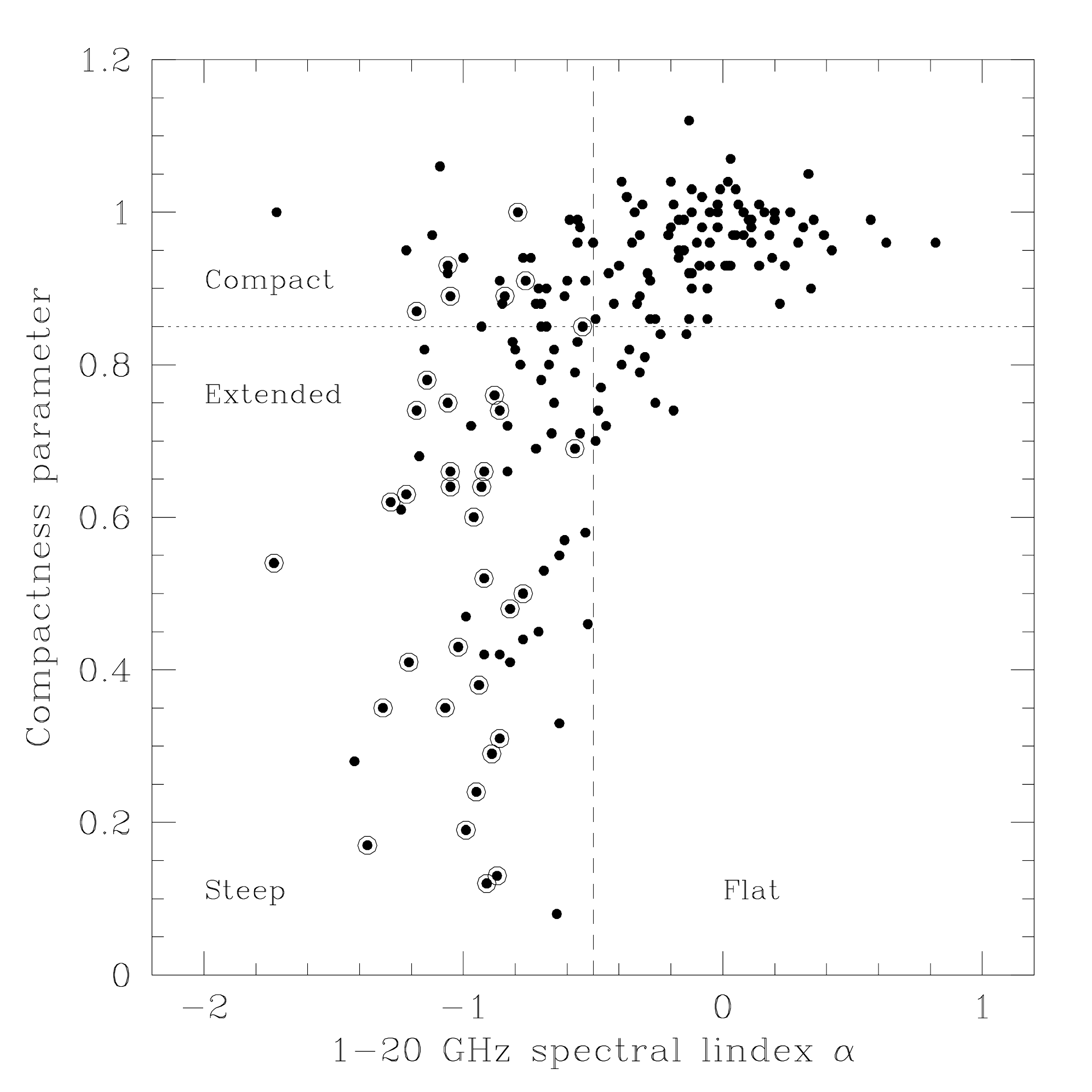}
\caption{Plot of the compactness parameter R (defined as the ratio of the flux densities measured on long and short baselines) at 20\,GHz against the 1-20\,GHz radio spectral index for AT20G-6dFGS galaxies, adapted from Sadler et al.\ (2014). Sources above the horizontal dotted line at R = 0.85 are expected to have angular sizes small\-er than about 0.2\,arcsec at 20\,GHz (i.e. smaller than about 220\,pc at the median redshift of $z=0.058$\ for this galaxy sample). Open circles show sources that are flagged as extended (on scales larger than 10-15 arcsec) in the AT20G catalogue, and the vertical dashed line at $\alpha=-0.5$\  shows the division between Ôsteep-spectrumÕ and Ôflat-spectrumÕ radio sources over the 1--20\,GHz frequency range. }
\label{fig_rajan}
\end{figure}

As can be seen from Figure\,\ref{fig_rajan}, almost half the AT20G-6dFGS sources are very compact at 20\,GHz based on the Chhetri et al.\ (2013) measurements, implying that their high-freq\-uency radio emission arises within the central 500\,pc (for a galaxy at $z\sim0.05$). Most of these very compact sources also have flat radio spectra over the frequency range 1--20\,GHz. The resolved AT20G-6dFGS sources, with steeper radio spectra, are mainly cores of radio galaxies with extended low-frequency emission as discussed below in \S2.4. 

\subsection{Are these beamed sources?} 
Some of the compact, flat-spectrum radio sources seen in the AT20G-6dFGS sample could be relativistically-beamed cores of FR-1 radio galaxies, where the central radio emission is boosted because a jet is viewed close to the line of sight, as is common in radio samples selected at high frequecy  (e.g. De Zotti et al.\ 2005). 

Sadler et al.\ (2014) looked at several different beaming diagnostics, including high-freq\-uency radio variability, the presence of broad optical emission lines, and evidence for a non-thermal continuum in the optical spectrum. They concluded that the fraction of flat-spectrum AT20G-6dFGS sources in which the observed radio emission was boosted by relativistic beaming was at most 35\% and could be significantly lower than this. In particular, these nearby compact sources do not generally show the high levels of radio variability characteristic of beamed sources. It therefore appears that most of the compact AT20G sources associated with nearby 6dFGS galaxies are candidate GPS/CSS sources rather than the beamed cores of low-luminosity radio galaxies.

\subsection{Source structure at lower frequencies} 
The AT20G-6dFGS sources represent a high frequency-sel\-ected sample of radio AGN, so our next step is to relate them to the more widely-studied samples of nearby radio AGN selected at 1.4\,GHz (e.g. Mauch \& Sadler 2007; Best \& Heckman 2012). The surface density of radio AGN in the AT20G-6dFGS sample (0.012\,deg$^{-2}$) is more than an order of magnitude lower than the value of  0.38\,deg$^{-2}$ for radio-loud AGN in the corresponding 1.4\,GHz-selected NVSS-6dFGS sample (Mauch \& Sadler 2007). This is because of the higher flux limit of the AT20G catalogue (40\,mJy at 20\,GHz) as well as the tendency for most radio AGN to be brighter at low frequencies.

Lower-frequency radio images of the AT20G-6dFGS are available at frequencies near 1\,GHz from the 1.4\,GHz NVSS and 843\,MHz SUMSS surveys, both of which have a $\sim45$\,arcsec FWHM beam (i.e. significantly lower spatial resolution than the high-frequency AT20G images).  The NVSS and SUMSS images do however have good sensitivity to  extended low-frequency emission from radio jets and lobes. 

Inspection of the SUMSS and NVSS images shows that most AT20G-6dFGS sources are also unresolved at lower frequency (see Table \,\ref{tab_structure}), with only 32\% classified as FR-1 or FR-2 radio galaxies on the basis of their low-frequency radio structure. 

%\pagebreak%%%%%%%%%%%%%%%%%

\begin{table}
% \centering%%%
\caption{Low-frequency (1\,GHz) radio structure of the AT20G-6dFGS sources. The `FR-0' sources (Ghisellini 2011) have angular sizes smaller than about 20\,arcsec at 1\,GHz, corresponding to a projected linear size of less than 15\,kpc at $z\sim0.05$; i.e.\ the low-frequency radio emission is likely to be entirely confined within the host galaxy. }
\label{tab_structure}
\begin{tabular}{lll}\hline
Structure & Fraction & Notes \\ 
%(unit1) & (unit2) \\
\hline
FR-2 & 8\% (16/201)   & Usually have double/triple \\
         &                         & structure at 20\,GHz \\
FR-1 & 24\% (49/201) & Usually have extended radio \\
          &                        & emission at 20\,GHz \\
FR-0  & 68\% (136/201) & Candidate GPS/CSS sources \\
\hline
\end{tabular}
\end{table}

\subsection{Compact radio AGN and the Fanaroff-Riley classification} 
The majority (68\%) of AT20G-6dFGS radio AGN fall into the class of compact `FR-0' radio sources defined by Ghisellini (2011), who noted that: {\it ``The `FR-0' radio ellipticals are a new population of radio sources (Baldi et al.\ 2009) having the same core radio luminosity as FR-1s, but hundreds of times less power in the extended emission.''} 

There is a close relationship between FR-0 radio AGN and GPS/CSS radio sources. Of the 136 `FR-0' radio AGN in Table\,\ref{tab_structure}, 25\% are candidate GPS sources with a radio spectrum peaking above 1--5\,GHz, 36\% are candidate CSS sources with a spectral peak below 5\,GHz, and the remaining 39\% are GPC/CSS candidates where there is not currently enough information to locate the spectral peak (in some cases these are AT20G sources north of declination -15$^\circ$, for which simultaneous 5 and 8\,GHz data are not available from the AT20G catalogue). 

\begin{figure}
\includegraphics[width=\linewidth]{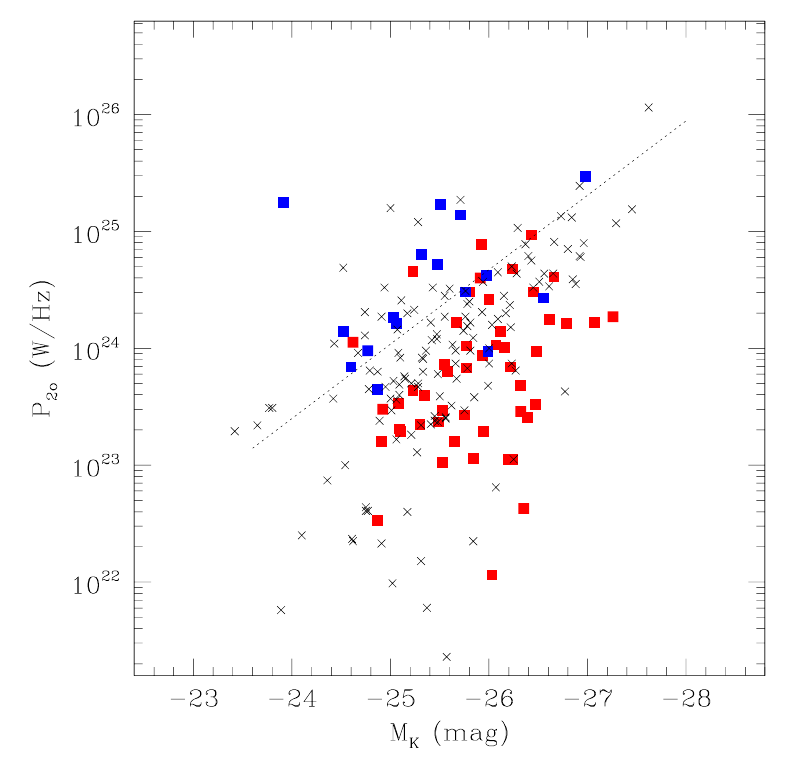}
\caption{Distribution of the AT20G-6dFGS galaxy sample in K-band absolute magnitude (which is closely related to galaxy stellar mass) and 20 GHz radio power. FR-1 radio galaxies are shown as filled red squares, FR-2 radio galaxies as filled blue squares and compact (FR-0) sources as black crosses. The dashed line corresponds to the 1.4 GHz FR-1/FR-2 dividing line from Ledlow \& Owen (1996), shifted to K-band and 20 GHz by assuming (i) a typical galaxy colour of (R-K) = 3.0\,mag. and (ii) a radio spectral index $\alpha=-0.74$ between 1 and 20\,GHz. }
\label{fig_lum}
\end{figure}

Figure\,\ref{fig_lum} shows the distribution of radio and optical luminosity for the AT20G-6dFGS galaxies.  A diagram of this kind was originally constructed at 1.4\,GHz by Ledlow and Owen (1996), who showed that the dividing line between FR-1 and FR-2 radio galaxies did not occur at a fixed radio luminosity but shifted to higher radio luminosity with increasing galaxy stellar mass. 

The reality of this dividing line has sometimes been questioned (e.g. Best et al.\ 2008; Gendre et al.\ 2013), but the 20\,GHz version shown in Figure\,\ref{fig_lum} uses the same volume-limited optical/radio sample to select all the objects plotted and so is largely free of selection effects.  It seems clear that  Ledlow and Owen (1996) dividing line, when shifted appropriately to higher frequency, represents the FR-1/FR-2 divide in the AT20G-6dFGS sample reasonably well. The `FR-0' radio AGN in the AT20G-6dFGS sample span at least three orders of magnitude in 20\,GHz radio power, and overlap in radio luminosity with both FR-1 and FR-2 sources hosted by galaxies of similar stellar mass.

\section{Global properties of the hosts of nearby radio AGN} 
Mid-infrared survey data from the all-sky Wide-field Infrared Survey Explorer (WISE) survey (Wright et al.\ 2010) can now be used to study the global properties of AGN host galaxies out to redshifts as high as $z\sim1$. 

\begin{figure}
\includegraphics[width=\linewidth]{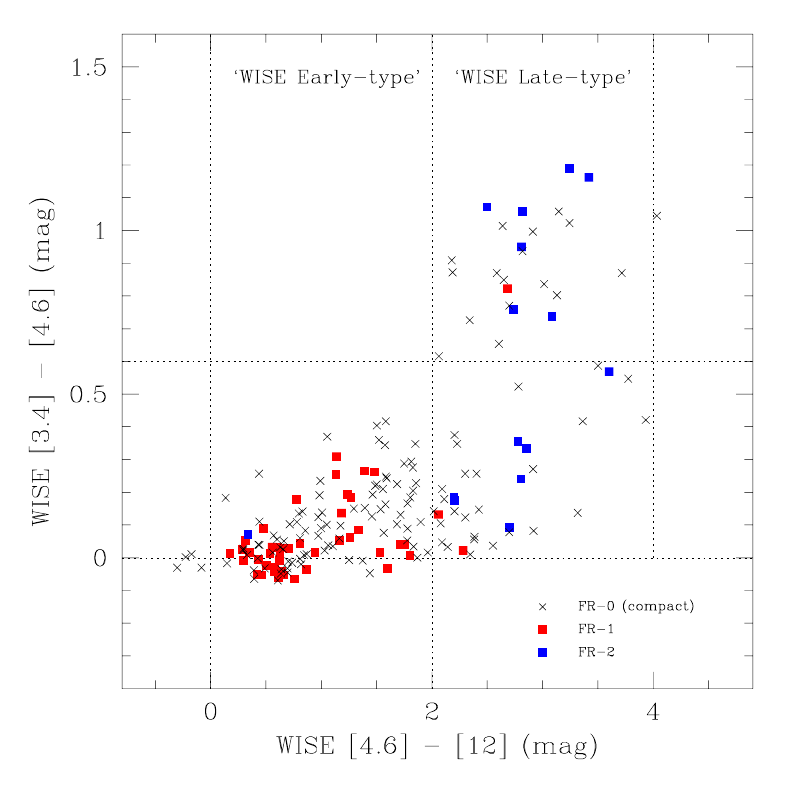}
\caption{ WISE colour-colour plot for for the host galaxies of FR-1 (red squares), FR-2 (blue squares) and compact (FR-0, black crosses) radio sources in the 20 GHz AT20G-6dFGS sample. The horizontal line at a $3.4-4.6\mu$m colour of 0.6 mag.\ divides the AGN and normal galaxy populations. Objects where radiation from an AGN dominates the galaxy spectrum in the mid-infrared are expected to lie above this line, and objects where starlight dominates should lie below the line.    }
\label{fig_wise}
\end{figure}

\subsection{The WISE two-colour plot} 
The WISE survey observed in four mid-IR bands, at 3.4, 4.6, 12 and 22$\mu$m.  Wright et al.\ (2010) have shown that a simple two-colour diagram like the one shown in Figure\,\ref{fig_wise}\ can distinguish between quiescent and star-forming galaxies, as well as between normal galaxies and those that host an AGN accretion disk. The WISE $4.6-12\mu$m\ colour is a useful first-order indicator of a galaxy's star-formation rate (Donoso et al.\ 2012), and can be used to distinguish early-type galaxies with a low star-formation rate from late-type galaxies where the star-formation rate is higher. Similarly, a WISE $3.4-4.6\mu$m\ colour with a value $>0.6$ indicates that a hot accretion disk is present in addition to starlight (Wright et al.\ 2010).

\subsection{WISE colours of the host galaxies of nearby radio AGN}
Figure\,\ref{fig_wise} shows a WISE two-colour plot for the host galaxies of radio AGN in the AT20G-6dFGS sample, split by FR classification.  
There is a strong split in the $4.6-12\mu$m\ colours of the host galaxies of FR-1 and FR-2 radio galaxies from the AT20G-6dFGS sample, with the FR-1 sources found almost exclusively in WISE Ôearly-typeÕ galaxies and the FR-2 sources (both HERGs and LERGs) in WISE Ôlate-typeÕ galaxies. 

This strongly suggests that the host galaxies of FR-1 and FR-2 radio sources are drawn from different galaxy populations, making it unlikely that an individual radio galaxy could evolve from an FR-2 to an FR-1 system.  Sadler et al. \ (2014) also find that the host galaxies of FR-1 radio sources in their sample are on average more massive than the hosts of FR-2 sources, consistent with earlier work (e.g. Baum, Zirbel \& O'Dea 1995)

It is interesting to note that the host galaxies of FR-0 sources span a wide range in WISE colours, with 67\% of FR-0 sources found in WISE early-type galaxies and 33\% in late-type galaxies with some ongoing star formation. This implies that the FR-0 sources, which make up the majority of the AT20G-6dFGS sample, are a more diverse population than the  FR-1 or FR-2 radio galaxies in the sample.

\section{What are the `FR-0' radio galaxies?} 
The compact `FR-0' sources which make up the majority of the AT20G-6dFGS sample are a heterogeneous population in terms of both their optical spectra (75\% LERGs, 25\% HERGs) and host galaxy type (67\% are in WISE early-type galaxies, 33\% in late-type systems). They also span a wide range in radio luminosity, from $10^{22}$ to $10^{26}$\,W\,Hz$^{-1}$\ at 1.4\,GHz. 
The overall radio properties of these FR-0 sources match those expected for young GPS and CSS radio sources, but a number of open questions remain to be resolved. 

In particular, the evolutionary path followed by FR-0 radio AGN remains unclear. The stellar mass of the host galaxy should remain roughly constant over the $10^7-10^8$\,yr 
lifetime of a typical radio AGN, so individual  points in Figure\,\ref{fig_lum} can shift up or down with time but cannot easily shift left or right. 

It therefore seems likely that some of the FR-0 sources in Figure\,\ref{fig_lum} could represent early stages in the evolution of FR-1 radio galaxies rather than the more powerful FR-2 objects. On the other hand, compact, low-luminosity radio AGN are far more common in the nearby Universe than would be expected if they all represented a short-lived early stage in the evolution of classical radio galaxies (e.g. Shabala et al. 2008; Turner \& Shabala 2015). Thus it seems likely that FR-0 sources will not evolve into large-scale FR-1/FR-2 radio galaxies. 

\begin{figure}
\includegraphics[width=\linewidth]{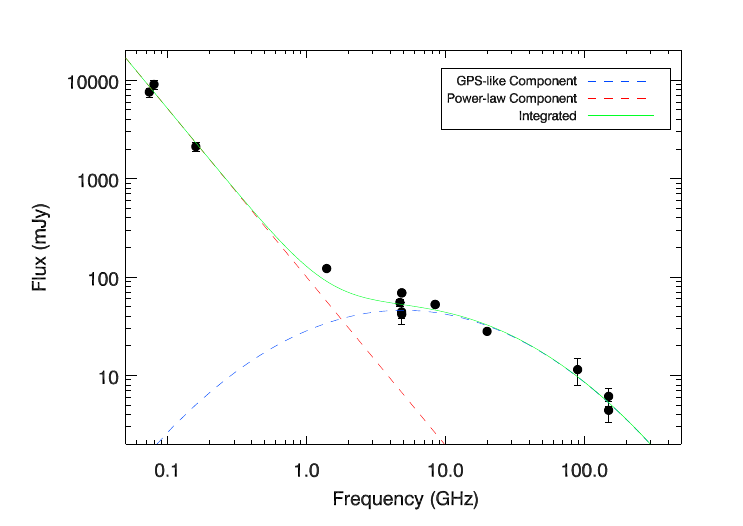}
\caption{Radio spectrum of the central galaxy in the nearby cluster Abell\ 496, from Hogan et al.\ (2015). This galaxy shows a high-frequency GPS-like component with a spectral peak near 10\,GHz, embedded in more diffuse low-frequency emission with a power-law radio spectrum. Objects of this kind can only be recognized by combining data across a wide range in radio frequency - in this case from 100\,MHz to 150\,GHz. }
\label{fig_hogan}
\end{figure}

One possibility is that FR-0 sources represent short-lived episodes of AGN activity which do not last long enough for a galaxy to develop large-scale radio jets. 
In this context, the results of the recent Hogan et al.\ (2015) study of nearby Brightest Cluster Galaxies (BCGs) are interesting. Many BCGs lie close to the cluster centre and can be fed by cooling gas from the halo, and the duty cycle of radio AGN in these objects is often close to 100\% (i.e. the radio core never completely switches off). Hogan et al.\ (2015) find that at least 3\% of BCGs (and at least 8\% of BCGs in cool-core clusters) have a radio spectrum with a GPS-like component which peaks above 2\,GHz. Figure\,\ref{fig_hogan}\ shows one example. 

A second possibility is that some FR-0 sources do have extended radio jets, but these jets have not been detected because they lie below the surface-brightness threshold of existing large-area radio surveys. If so, the new low-frequency (100-200\,MHz) radio surveys now being carried by LOFAR and MWA may detect some of these faint jets. The recent MWA detection of diffuse, large-scale radio jets in the nearby galaxy NGC\,1534 (Hurley-Walker et al.\ 2015) is one example where low-frequency observations reveal previously-undetected evidence of AGN activity. 

An additional possibility, discussed in the presentation by Baldi at this meeting (Baldi, Capetti \& Giovannini 2015b), is that the radio jets in FR-0 objects are slow, and more easily-disrupted radio jets than the jets in FR-1/FR-2 radio galaxies, because the central black hole spin is too slow to power a fast jet. Although black-hole spin is difficult to measure observationally, there may be subtle differences in the environment and/or dynamics of FR-0 and FR-1 host galaxies which reflect a different merger history (Baldi et al.\ 2015a).

\section{Summary} 
As a summary, I return to the two questions posed in \S1.3: 
 
{\it 1. Is there a large population of  lower-luminosity (radio power  $<10^{25}$\,W\,Hz$^{-1}$) GPS/CSS sources? } \\
While most well-studied GPS and CSS sources are powerful (typically $>10^{25}$\,W\,Hz$^{-1}$ at 1.4\,GHz) radio sources with a compact double morphology, I think the focus on these bright objects comes about at least in part because of observational selection effects. It is clear that lower-lum\-inosity GPS and CSS sources do exist (as also discussed in the presentation by Kunert-Bajraszewska at this meeting (Kunert-Bajraszewska 2015)). Such sources are probably quite numerous, but they have been difficult both to identify (because of a lack of sensitive large-area surveys at multiple radio frequencies) and to study in detail (because of the time needed to carry out detailed VLBI imaging of fainter radio sources). 

{\it 2. If so, how do their properties compare to the more luminous GPS and CSS samples studied to date? } \\
Although recent observations reveal the existence of a significant population of compact, lower-luminosity radio sources (as also discussed by Baldi at this meeting; Baldi et al.\ 2015b), their relationship to the classical GPS and CSS objects remains unclear. Much work still remains to be done, both observationally (e.g. more detailed VLBI studies) and in developing demographic and evolutionary models for this low-luminosity population. In some cases, the presence of a GPS-like component only becomes evident at frequencies above 5-10\,GHz, while observations at frequencies well below 1\,GHz may be needed to reveal the presence of `relic' emission from extended radio jets.  Future wide-band radio continuum surveys, coupled with matching optical/infrared data, are needed to advance our understanding of the compact radio-source population, and these surveys should include the widest possible coverage in radio frequency - ideally from 100\,MHz to 100\,GHz! \\

\acknowledgements
I thank the organisers of the Rimini GPS/CSS Workshop for an extremely pleasant, stimulating and informative meeting. I am also grateful to many colleagues, 
including James Allison, Scott Croom, Alastair Edge, Ron Ekers, Mike Hogan, Elizabeth Mahony, Tom Mauch, Tara Murphy and Steven Tingay, for useful discussions and for allowing me to include some of their recent results in my presentation.


\begin{thebibliography}{}
 \bibitem{} Baldi, R. D., Capetti, A.: 2009, A\&A~508, 603
 \bibitem{} Baldi, R. D., Capetti, A., Giovannini, G.: 2015a, A\&A~576, 38
 \bibitem{} Baldi, R. D., Capetti, A., Giovannini, G.: 2015b, AN, in press 
 \bibitem{} Baum, S.A., Zirbel, E.L., O'Dea, C.P.: 1995, ApJ~451, 88 
 \bibitem{} Best P. N., Kauffmann G.. Heckman, T. M., Brinchmann, J., Charlot, S.. Ivezic, Z., White, S. D. M.: 2005, MNRAS~362, 25 
 \bibitem{} Best, P. N., Heckman, T. M.: 2012, MNRAS~421, 1569 
 \bibitem{} Chhetri, R., Ekers, R. D., Jones, P. A., Ricci, R.: 2013, MNRAS~434, 956 
 \bibitem{} Cohen, A. S. et al.: 2007, AJ~134, 1245-1262 
 \bibitem{} Condon, J. J.: 1989, ApJ~338,13
 \bibitem{} Condon, J. J., Cotton, W. D., Greisen, E. W., Yin, Q. F., Perley, R. A., Taylor, G. B., Broderick, J. J.: 1998, AJ~115, 1693
 \bibitem{} de Vries, N., Snellen, I. A. G., Schilizzi, R. T., Lehnert, M. D., Bremer, M. N.: 2007, A\&A~464, 879 
 \bibitem{} De Zotti G., Ricci R., Mesa D., Silva L., Mazzotta P., Toffolatti L., Gonzalez--Nuevo J.: 2005, A\&A~431, 893
 \bibitem{} Fanaroff, B. L., Riley, J. M.: 1974, MNRAS~167, 31 
 \bibitem{} Fosbury, R. A. E., Bird, M. C., Nicholson, W., Wall, J. V.: 1987, MNRAS~225, 761
 \bibitem{} Ghisellini, G., 2011: American Institute of Physics Conference Series~1381, 180
 \bibitem{} Gregory, P. C., Vavasour, J. D., Scott, W. K., Condon, J. J.: 1994, ApJS~90, 173 
 \bibitem{} Gregory, P. C., Scott, W. K., Douglas, K., Condon J. J.: 1996, ApJS~103, 427 
 \bibitem{} Hancock, P. J., 2009: AN~330, 180
 \bibitem{} Hardcastle, M. J., Evans, D. A., Croston, J. H.: 2007, MNRAS~376, 1849 
 \bibitem{} Heckman, T. M., Best, P. N.: 2014, ARAA~52, 589 
 \bibitem{} Hogan, M. T. et al.: 2015, MNRAS in press, arXiv 1507.03022
 \bibitem{} Hurley-Walker, N. et al.: 2015, MNRAS~447, 2468 
 \bibitem{} Jones, D. H. et al.: 2009, MNRAS~399, 683 
 \bibitem{} Kunert-Bajraszewska, M., Gawronski,M.P., Labiano, A., Siemiginowska, A.: 2010, MNRAS~408, 2261
 \bibitem{} Kunert-Bajraszewska, M.: 2015, AN, in press
 \bibitem{} Labiano, A., Barthel, P. D., OÕDea, C. P., de Vries, W. H., Perez, I., Baum, S. A.: 2007, A\&A~463, 97
 \bibitem{} Ledlow M. J., Owen F. N.: 1996, AJ~112, 9 
 \bibitem{} Mauch T. et al.: 2003, MNRAS~342, 1117
 \bibitem{} Mauch T., Sadler E. M., 2007: MNRAS~375, 931 
 \bibitem{} Murphy, T. et al., 2010: MNRAS~402, 2403 
 \bibitem{} O'Dea, C. P.: 1998, PASP~110, 493
 \bibitem{} O'Dea, C. P., Baum, S. A., Stanghellini, C.: 1991, ApJ~380,66 
 \bibitem{} Randall, K. E., Hopkins, A. M., Norris, R. P., Edwards, P. G.: 2011, MNRAS~416, 1135
 \bibitem{} Rengelink, R. B. et al.: 1997, A\&AS~124, 259 
 \bibitem{} Reynolds, J.: 1994, ATNF Technical Memo AT/39.3/040 (see http://www.atnf.csiro.au/observers/memos/)
 \bibitem{} Sadler E. M.\ et al.: 2002, MNRAS~329, 227
 \bibitem{} Sadler, E. M., Ricci, R.,Ekers, R. D., Sault, R. J., Jackson, C. A., De Zotti, G.: 2008, MNRAS~385, 1656
 \bibitem{} Sadler, E. M., Ekers, R. D., Mahony, E. K., Mauch, T., Murphy, T.: 2014, MNRAS~438, 796
 \bibitem{} Shabala, S. S., Ash, S., Alexander, P., Riley, J. M.: 2008, MNRAS~388, 625 
 \bibitem{} Snellen, I. A. G., Mack, K.-H., Schilizzi, R. T., Tschager, W., 2003:  PASA~20, 38 
 \bibitem{} Tingay, S. J., Edwards, P. G., Tzioumis, A. K.: 2003, MNRAS~346, 327 
 \bibitem{} Tingay, S. J., Edwards, P. G.: 2015, MNRAS~448, 252 
 \bibitem{} Tsioumis, A. K.: 2010, AJ~140, 1506 
 \bibitem{} Turner, R. J., Shabala, S. S., 2015: ApJ~806, 59
 \bibitem{} Urry, C. M., Padovani, P. : 1995, PASP~107, 803 
 \bibitem{} van Velzen, S., Falcke, H., Schellart, P., Nierstenh\"ofer, N., Kampert, K.-H.: 2012, A\&A~544, A18
 \bibitem{} Wright, E. L. et al.: 2010, AJ~140, 1868
 \end{thebibliography}
\end{document}